\documentclass[aps,pra,showpacs,twocolumn,superscriptaddress]{revtex4-1}
\usepackage{graphicx}
\usepackage{dcolumn}
\usepackage{bm}
\usepackage{amssymb}
\usepackage{amsmath}
\usepackage{epsfig}
\usepackage{hyperref}
\usepackage{times}

\newcommand{\ket}[1]{\left| #1 \right\rangle}
\newcommand{\bra}[1]{\left\langle #1 \right|}

\newcommand{\pder}[2]{\frac{\partial#1}{\partial#2}}

\input{tcilatex}

\begin{document}

\title{Advantages of Multi-photon Detection Revealed by Fisher Information In Resolving Incoherent Sources}
\author{Xiao Liang}
\author{Yong-Sheng Zhang}
\email{yshzhang@ustc.edu.cn}
\affiliation{Laboratory of Quantum Information, University of Science and Technology of China, Hefei, 230026, China}
\affiliation{Synergetic Innovation Center of Quantum Information and Quantum Physics, University of Science and Technology of China, Hefei, 230026, China}
\date{\today }

\begin{abstract}
With the progress of optical detection technology, the classical diffraction limit raised a hundred years ago has been continuously broken through. In previous experiments within fluorescence sources, one of the techniques used is detecting auto-correlation functions. The image is reconstructed by single-photon and multi-photon detecting intensities. By taking the joint intensity of more than one single-photon detectors into consideration, the point spread function of each photon emitter can be resolved even when the distance of their central position is below the classical diffraction limit. However, the measurement precision is not considered. In actual detecting process the detectors are imperfect, they have quantum detecting efficiency $\eta$ and even can not count the photon numbers during one absorbing process. Therefore, we analyze the detecting intensity of each detector separately and use the total Fisher information to depict the resolution. Higher Fisher information is obtained when second order correlation function is considered and higher detecting efficiency is beneficial to the Fisher information enhancement, meanwhile the Cram\`er-Rao bound for an effective distance resolving is lower than the classical diffraction limits. Furthermore, when the emitted photons are coherent states, the resolution ability can be enhanced with infinite single-photon detectors.
\end{abstract}

\maketitle
\section{Introduction}
The resolution of optical imaging is a vital benchmark in physics and many technology applications. More than a century ago, the lowest bound that the two incoherent light spots can be resolved is known as the classical diffraction limit \cite{1} or the Rayleigh criterion \cite{2}. Since improving resolution in optical imaging is significant in astronomy, earth science, medical research and many other scientific fields, superresolution beyond the Rayleigh criterion via technical methods has attracted more and more interests in recent years.

Two primary ideas in achieving superresolution of two sources are: (1) stimulated-emission-depletion \cite{3,4}; (2) multi-photon detection, such as the large repeat times \cite{5}, quantum entanglement \cite{6} and the correlation function measurements \cite{7,8}. In these experiments, the objects being imaged are usually fluorescence sources, such as nitrogen-vacancy center (NVC) \cite{9}, single molecule \cite{5} and biological cell \cite{10}. Recently, some theoretical work on achieving the resolution beyond the Rayleigh criterion are focused on the aspects such as the construction of quantum basis used for single-photon detection \cite{11}, quantum entanglement assisted superresolution \cite{12,13} and the detection of correlation functions \cite{14}. These works gave new perspectives on how quantum mechanics can beat the classical diffraction limits and support the experimental methods used on achieving superresolution. Because of photon antibunching effect, each source emits no more than one photon during an emission process \cite{15} in these systems. The high order correlation function can effectively depict the overlapped area between the sources. When the emitted light from the sources is coherented, the resolving ability can be enhanced by the factor of $N$ with the help of detecting $N$-photon entangled states \cite{16}. With incoherent light sources, the intensity of first order detecting is the summation of the each light's intensity \cite{11,12}. However, to achieve the superresolution the first order detecting intensity is not enough, more information can be extracted from the joint intensity of two single-photon detectors. Such joint intensity is revealed by the second order correlation function of the photons and can effective depict the overlapped area between the point spread functions (PSFs) of the emitters. 

The resolution limit of the distance between the PSFs is determined by the detecting intensity of the detectors. In real cases such as resolving two Airy disks, each point spread function (PSF) is obtained based on the detecting intensities of the detectors. For ideal detectors, the normalized single-photon detecting intensity is: $I(x)=\frac{1}{2}\left[I_A(x)+I_B(x)\right]$, where $I_A$ and $I_B$ denote the intensities of two incoherent light emitters. A small displacement between the PSFs causes a variance of the intensity: $I(x|d)-I(x|0)$. Based on the definition of Fisher information, the variance of the intensity about to distance is indicated by the intensity change rate summarized at all pixels \cite{17}: 
\begin{equation}
F_{d}=\sum_{x}\frac{1}{I(x|d)}\left[\pder{I(x|d)}{d}\right]^2.
\label{id}
\end{equation}

The resolution ability can be indicated by the Fisher information because it is easier to distinguish the PSFs with rapider intensity variations, and the corresponding Cram\`er-Rao bound is the lowest measuring error of physical variables. Therefore, the Fisher information reveals the resolution ability from the perspective of the shape of the distributions, and the Cram\`er-Rao bound reveals the resolution ability of the measurements of physical variables. However without considering the detecting precision, the distributions of the PSFs can still be obtained experimentally \cite{9,18,19}.

Above all, the measurement precision of superresolution should be investigated, especially with imperfect single-photon detectors. How will the multi-photon detection advantages with imperfect single-photon detectors? Can the classical diffraction limit still be beaten with imperfect single-photon detectores? How to amplify the enhancement of multi-photon detection with imperfect single-photon detectors?  Here, we analyzed the Fisher information based on the detecting intensities of the exclusive events and obtain the Cram\`er-Rao bound. When resolving the PSFs from two single-photon emitters, at least two single-photon detectors should be used, we focus on the exclusive events that: 1, only one of the detectors clicks; 2, both detectors click simultaneously. In addition, we propose that more single-photon detectors are beneficial to resolution with multi-photon emitters. 

This paper is organized as follows: In Sec.~\ref{sec:A} we derive the detecting probability of the exclusive events with two single-photon emitters and two single-photon detectors. In Sec.~\ref{sec:B} we calculate the total Fisher Information based on the exclusive events and obtain the Cram\`er-Rao bound of the distance measurements. In Sec.~\ref{sec:C} we analyze the detecting intensities with $n$ single-photon detectors with coherent-state photon emitters. In Sec.~\ref{sec:D} we summarize our work and draw our conclusions.

\section{the exclusive detecting events with two single-photon emitters}
\label{sec:A}
\begin{figure}
\includegraphics[width=0.5\textwidth]{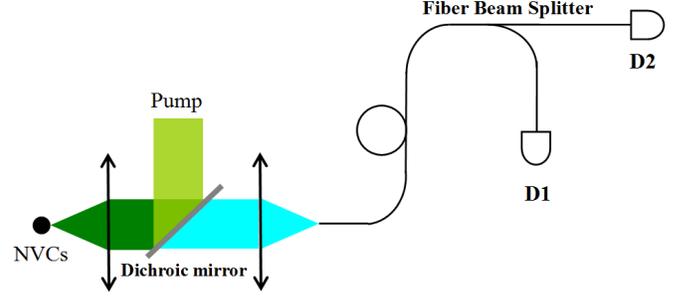}
\caption{ (color online) The scanning confocal system used in the experiments. Photons emitted from NVCs are collected into the fiber and are split into two paths by a fiber beam splitter. The fiber and the single-photon detectors form a Hanbury Brown-Twiss interferometer. }
\label{fig1}
\end{figure}

We consider the sources that are NVCs pumped by a scanning confocal system \cite{20}. The imaging experiment setup is depicted in Fig.(\ref{fig1}), photons emitted from NVCs are focused by optical lens, then collected into a single-mode fiber and split into two paths by a fiber beam splitter that forms a Hanbury Brown-Twiss interferometer \cite{21}. To form the entire image, the NVCs' position is moved relative to the scanning confocal system.  For a single NVC, there is at most one photon emission at any time, the state of the photon collected by the fiber at position $x$ is: $\rho_k(x)=P_k(x)\ket{1}\bra{1}+\left[1-P_k(x)\right]\ket{0}\bra{0}$, where $k$ denotes the photon is emitted from the $k$-th source and $P_k(x)$ is the point spread function. Since the NVCs are uncorrelated, the density of the emitted photons is: $\rho=\mathop{\otimes}\limits_{n}\rho_k$ \cite{7}. The measurement outcomes are indicated by the clicks of the detectors. When single-photon detection is performed, the absorption operator of the total electric field is the summation from both sources: $\hat{E}^{(+)}(r)=\xi\left(\hat{a}_Ae^{ik_Ar}+\hat{a}_Be^{i\phi}e^{ik_Br}\right)$, $k_A$ and $k_B$ are the wave-vectors of the photons emitted from source A and B, $r$ is the position of the electric field and $\phi$ is the random phase between the emitters. We assume that the beam splitter is semi-reflected and semi-transmitted and denote the path ends at detector $D1$ as path $L$ and the path ends at $D2$ as path $R$. The absorption operators of the electric field under the effect of the beam splitter becomes $\hat{E}^{(+)}(x)=\frac{\xi}{\sqrt{2}}\left(\hat{a}_{A}e^{ik_A r_R}+i\hat{a}_{A}e^{ik_A r_L}+\hat{a}_{B}e^{i\phi}e^{ik_B r_R}+i\hat{a}_{B}e^{i\phi}e^{ik_B r_L}\right)$, the symbols $A$ and $B$ denote the photon is emitted from source $A$ and source $B$, $L$ and $R$ denote the photon is on path $L$ and path $R$, respectively. Here, we care about the situation that at the same time $D1$ clicks but $D2$ does not click. For studying the clicks of the detectors, we focus on the electric field on the paths where detector exists. From the expansion of the total electric field, the electric field operators at path $L$ and path $R$ are: $\hat{a}_L=i\frac{\xi}{\sqrt{2}}\left(\hat{a}_{A}e^{ik_A r_L}+\hat{a}_{B}e^{i\phi}e^{ik_B r_L}\right)$ and $\hat{a}_R=\frac{\xi}{\sqrt{2}}\left(\hat{a}_{A}e^{ik_A r_R}+\hat{a}_{B}e^{i\phi}e^{ik_B r_R}\right)$. Therefore the single-photon intensity at path $L$ is: 
\begin{equation}
\text{Tr} \rho \hat{a}_{L}^\dag\hat{a}_{L}=\frac{1}{2}\left[P_A(x)+P_B(x)\right], 
\label{1}
\end{equation}
where $P_A(x)$ and $P_B(x)$ are the PSFs of the emitters. The total light intensity is proportional to detection intensity only when one photon is emitted, such conclusion is mentioned in Ref \cite{9}. Next we focus on the situation when two photons are collected, the probability for both path $L$ and path $R$ has one photon is revealed by the correlation function:
\begin{equation}
\text{Tr}(\rho\hat{a}_L^\dag\hat{a}_R^\dag\hat{a}_L\hat{a}_R)=P_A(x)P_B(x).
\label{2}
\end{equation}

Since the photons emitted from source $A$ and $B$ are uncorrelated. On path $L$, the probability of the photon emitted from source $A$ is: $\text{Tr}(\rho_A\otimes I\hat{a}_L^\dag\hat{a}_L)=\frac{1}{2}P_A(x)$. And the probability of the photon emitted from source $B$ is $\text{Tr}(I\otimes\rho_B\hat{a}_L^\dag\hat{a}_L)=\frac{1}{2}P_B(x)$. Because of the uncorrelated photon emitters, the joint probability for two photon in path $L$ is the multiplication of the probabilities of photon $A$ and $B$: 
\begin{equation}
\frac{1}{4}P_A(x)P_B(x).
\label{3}
\end{equation}

The probability for both photons in path $R$ is the same due to 50:50 beam splitter. When the detectors are not perfect, a photon at the corresponding path has the probability of $\eta$ to be absorbed and the probability of $(1-\eta)$ to be unabsorbed. Suppose that the detectors can not distinguish photon number, they can only distinguish the cases of zero and non-zero photon absorption. Therefore, combining all of the probabilities above we have the probability of event $D1$ clicks while $D2$ does not click, we name such event as $\alpha$. The detecting intensity of event $\alpha$ is:
\begin{equation}
\begin{split}
P_\alpha(x)=&\frac{1}{2}\eta\left[P_A(x)+P_B(x)\right]+\\
&P_A(x)P_B(x)\left\{\eta(1-\eta)+\frac{1}{4}\left[1-(1-\eta)^2\right]\right\},
\label{4}
\end{split}
\end{equation}
where $P_\alpha(x)$ consists of the situations when only one photon is collected (Eq.\ref{1}), two photons are simultaneously collected but each photon is on different path (Eq.\ref{2}) and both photons are on the same path (Eq.\ref{3}).
Since the 50:50 beam splitter, the event $\beta$ that $D2$ clicks while $D1$ does not click has the same probability as $\alpha$. Based on the results in Eq.(\ref{2}), the detecting intensity that $D1$ and $D2$ click simultaneously is:
\begin{equation}
P_\gamma=\eta^2P_A(x)P_B(x).
\label{5}
\end{equation}

Combine with Eq.(\ref{4}) and Eq.(\ref{5}), we can calculate the total Fisher information of the joint events. For obtaining the Fisher information of the experiment setups, we divided the experiment outcomes into three exclusive events: $\alpha$, $\beta$ and $\gamma$. Each event is normalized in all of the exclusive events:
\begin{equation}
C_m(x)|_{m=\alpha,\beta,\gamma}=\frac{P_m(x)|_{\alpha,\beta,\gamma}}{\sum_{m=\alpha,\beta,\gamma}\int_{-\infty}^{\infty}P_m(x)dx}.
\label{6}
\end{equation}

\section{The Fisher Information of Exclusive Detecting Events}
\label{sec:B}
\begin{figure}
\includegraphics[width=0.5\textwidth]{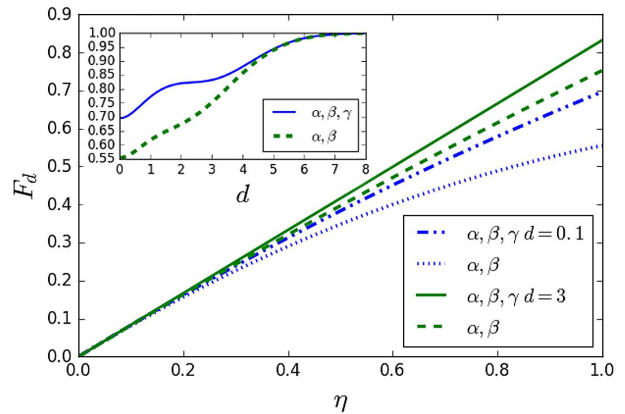}
\caption{ (color online) The solid green line are the Fisher information about variable $\eta$ when events $\alpha$, $\beta$ and $\gamma$ are considered when $d=3$. The dashed green line depicts the Fisher information when only events $\alpha$ and $\beta$ are considered. The dash-dotted blue line is $d=0.1$ with $\alpha$, $\beta$ and $\gamma$ are considered. The dotted blue line is the Fisher information with $\alpha$ and $\beta$. The sub figure reveals the relation between Fisher information and the distance according to combinations of the events when the detecting efficiency $\eta=1$. Here we chose $\sigma=1$ and $M=1$. }
\label{fig2}
\end{figure}

Here we consider two detectors, therefore there are at most three exclusive events as mentioned in the last section: $\alpha$, $\beta$ and $\gamma$. According to the Fisher information additivity of the independent incidents, the total Fisher information is the summation of each event \cite{12,22}:
\begin{equation}
F_d=N_{eff}f_d=N_{eff}\sum_{m=\alpha,\beta,\gamma}\int\limits_{-\infty}^{\infty} C_m(x)\left[\frac{\partial\text{ln}C_m(x)}{\partial d}\right]^2 dx,
\label{7}
\end{equation}
where $f_d$ is the Fisher information of the normalized distributions in Eq.(\ref{6}). $N_{eff}$ is the effective repeating times of the imaging process and $N_{eff}=M\sum\limits_{m}\int\limits_{-\infty}^{\infty}P_m(x)dx$, $M$ is the repeating times of the imaging process and $\sum\limits_{m}\int\limits_{-\infty}^{\infty}P_m(x)dx$ is the photon flux density in unit time. Here we assume that the PSFs are Gaussian distributed with their central positions have distance $d$: $P_A(x)=\frac{1}{\sqrt{2\pi}\sigma}\exp(-\frac{x^2}{2\sigma^2})$ and $P_B(x)=\frac{1}{\sqrt{2\pi}\sigma}\exp\left[-\frac{(x-d)^2}{2\sigma^2}\right]$ \cite{12}. The variable to be measured is the distance value $d$. Submitting Eq.(\ref{4}) and Eq.(\ref{5}) into Eq.(\ref{6}), based on the fact that $P_\alpha(x)=P_\beta(x)$, and using the Gaussian assumption and Eq.(\ref{7}), the Fisher information is numerical calculated. It is clearly shown in Fig.(\ref{fig2}) that larger distance results higher Fisher information. Because it is the easier to resolve the PSFs with wider distance, and the detecting uncertainty of two well-resolved PSFs is low. However when the event $\gamma$ is taken into consideration, more information can be obtained from the over-lapped area of the PSFs. Therefore the multi-photon detection can enhance the Fisher information with all the detecting efficiencies.  When $\eta\rightarrow0$, no information can be obtained as $F_d\rightarrow0$. The Fisher information enhancement caused by the second order detecting intensity takes effect with small distances, as the sub figure in Fig.(\ref{fig2}) depicts. When the PSFs are well seperated, they can already be resolved by the detecting intensity of one detector and it is not necessary to take the second order detecting intensity into consideration.
\begin{figure}
\includegraphics[width=0.5\textwidth]{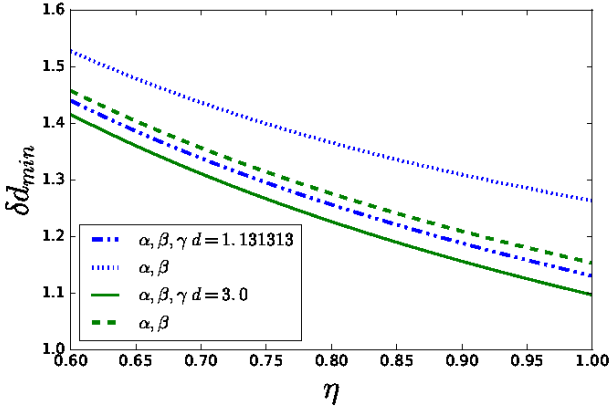}
\caption{ (color online) The relation between the Cram\`er-Rao bounds and single-photon detecting efficiensie. The green solid line depicts the Cram\`er-Rao bound of events $\alpha$. $\beta$ and $\gamma$, and the green dashed line depicts the Cram\`er-Rao bound of events $\alpha$ and $\beta$. At the critical distance $d=1.131313$, as the blue dotted line depicts, the Cram\`er-Rao bound is larger than signal under all of the detecting efficiensies when $\gamma$ is not considered. However the Cram\`er-Rao bound is the same as the signal combining events $\alpha$, $\beta$ and $\gamma$ with perfect single-photon detecting efficiency , as depicted by the blue dot-dashed line.  Here we chose $\sigma=1$ and $M=1$. }
\label{fig3}
\end{figure}

In the distance measurements, we consider the Cram\`er-Rao bound of the measurement error which is determined by one effective imaging process \cite{12,23}. Here we consider one imaging process ($M=1$). To achieve the low Cram\`er-Rao bound we can independently repeat the imaging process by many times. As we can have a gain of the Fisher information of $M$ times, based on the relations between the detecting error and the Fisher information: $\delta d\geq\frac{1}{\sqrt{F_d}}$. The Cram\`er-Rao bound can be diminished by: $\frac{1}{\sqrt{M}}$. However in one imaging process, the Cram\`er-Rao bound is the least measuring error by whatever techniques that can be performed. To effectively decrease the measuring error by large independent repeating times, it is required that in each imaging process the Cram\`er-Rao bound is lower than the signal itself. Otherwise, the uncertainty of each imaging will be accumulated in the independent repeating of the imaging process, and the obtained value is inaccurate.

Based on the Fisher information in Fig.(\ref{fig2}), the Cram\`er-Rao bound is depicted in Fig.(\ref{fig3}). It is shown that for a fixed distribution of the PSFs and a certain mixture of detecting events, the detecting error is lower with better detecting efficiencies. And the consideration of second order detecting intensity $\gamma$ will enhance the Fisher information, that is, to decrease the measuring error. In a single imaging process with perfect detectors ($\eta=1$), the critical distance that can be measured by events $\alpha$, $\beta$ and $\gamma$ but not obtainable by events $\alpha$ and $\beta$ is approximately $d=1.131313$, which is below the classical diffraction limit. In the profile of a Gaussian approximation to an Airy disk, the radius of the Airy disk is about three times as the width of the Gaussian distribution \cite{24}. Therefore, the consideration of second order correlation  can effectively increase the resolving abilities. 

\section{The $n$-th order detecting intensity of Multi-photon sources}
\label{sec:C}
\begin{figure}
\includegraphics[width=0.5\textwidth]{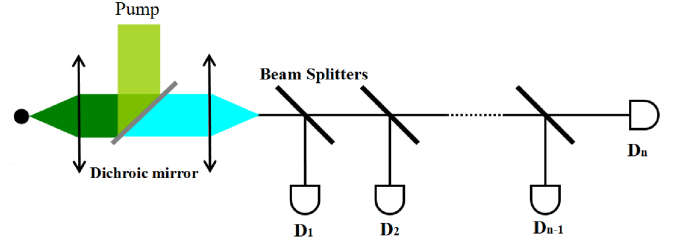}
\caption{ (color online) The experiment setup of $n$ single-photon detectors and $(n-1)$ beam splitters. The photons emitted from sources are coherent states and are collected by the optical fiber. After entering the fiber, the photons are seperated to $n$ single-photon detectors by $(n-1)$ beam splitters. }
\label{fig4}
\end{figure}

Because of photon anti-bunching effect, there are at most two photons emitted from two NVCs. However when the emitted photons are coherent states $\ket{z}$ such as lasers, the number of the photons that can be simultaneously detected is infinite. Therefore, $n$-th order detection can be performed with $n$ single-photon detectors and $(n-1)$ beam splitters, where $n$ can be a very large number. The corresponding experiment setup is depicted in Fig.(\ref{fig4}).

Because of the un-correlated photon emitters, the system density of the emitted photons is: $\rho=\mathop{\otimes}\limits_{n}\rho_i$, where $\rho_i$ is revealed by the PSF of the coherent light: $\rho_{i}=P_{i}(x)\ket{z}\bra{z}$. Here we assume that all beam splitters are half reflected and half tranmitted. Therefore, the electric field at the $k$-th detector is: $\hat{a}_k=\frac{\xi}{2^{k/2}}(\hat{a}_Ae^{ik_Ar_k}+\hat{a}_Be^{i\phi}e^{ik_Br_k})$, except for the last detector: $\hat{a}_n=\frac{\xi}{2^{(n-1)/2}}(\hat{a}_Ae^{ik_Ar_n}+\hat{a}_Be^{i\phi}e^{ik_Br_n})$, where $kr$ is the propagation phase and $\phi$ is the random phase between the emitters. The $n$-th order detecting intensity is revealed by the circumstance when $n$ detectors click simultaneously:
\begin{equation}
P_\gamma=\text{Tr}\left(\rho\hat{a}_1^\dag\hat{a}_2^\dag\cdots\hat{a}_n^\dag\hat{a}_1\hat{a}_2\cdots\hat{a}_n\right).
\label{8}
\end{equation}  

Since the paths of the photons are limited by the optical fiber, the spreading phase equals to the scalar product of wave-vector and distance. For simplicity we assume that light emitted from two sources have the same wavelength. Therefore, Eq.(\ref{8}) can be rewritten as:
\begin{equation}
P_\gamma=2^{(1-n)(n+2)/4}{|\xi|}^{2n}\text{Tr}\rho\sum_{k=0}^{n}\binom{n}{k}^2(\hat{a}_A^\dag)^k\hat{a}_A^k(\hat{a}_B^\dag)^{n-k}{\hat{a}_B}^{n-k}.
\label{9}
\end{equation}

Since there are two photon sources in our experiment setup, the intensity of the PSFs we measure are the combination of first-order light intensity: $P_A(x)+P_B(x)$ and the intensity of overlapped area: $P_A(x)P_B(x)$. It is concluded in the last section that overlapped area contains more information of the distance between the central positions of the PSFs. The distribution proportion of over-lapped area and first-order detecting intensity in $n$-th order detecting intensity can be calculated from Eq.(\ref{9}):
\begin{equation}
\frac{\text{Over-lapped Intensity}}{\text{First-order Intensity}}=\frac{\sum_{k=1}^{n-1}\binom{n}{k}^2P_A(x)P_B(x)}{P_A(x)+P_B(x)}.
\label{10}
\end{equation}

It is revealed in Eq.(\ref{10}) that the proportion of second-order correlation increases with more single-photon detectors. With infinite single-photon detectors, the $n$-th order detecting intensity is exactly the over-lapped area of the PSFs. Therefore, the detecting precision can be enhanced with more single-photon detectors when the emitted photons are coherent states. 

From the calculation in Eq.(\ref{9}) we can find that the $n$-th order detecting intensity is proportional to $|z|^{2n}$, where $|z|^2$ is the mean photon number of the coherent states. With weak coherent light ($z\ll1$), the coherent state is expanded to first order: $\ket{0}+z\ket{1}$, therefore weak coherent light is similar to a single-photon emission with approximated probability ${|z|}^2$, meanwhile the $n$-th order detecting intensity is weak. However when the light intensity is strong ($z\gg1$), the incident light is equivalent to classical light, the $n$-th order detecting intensity dominates. Therefore, the multi-photon detection still have advantages with classical light sources because of high detecting probability. What's more, the Fisher Information of multi-photon detection can be enhanced by using more single-photon detectors. 

\section{Summary and Conclusion}
\label{sec:D}
The classical diffraction limit has been continuously discussed that whether it is the real bound in resolution or it is just an experimental parameter and can be beaten via quantum mechanics. Nowadays, the classical diffraction limit (or the Rayleigh criterion) is commonly known as an experimental parameter. The Cam\`er-Rao bound determined by the detecting device is investigated in the article. As depicted by Fig.(\ref{fig2}) and Fig.(\ref{fig3}), the enhancement of Fisher information is mainly achieved by the usage of multi-detectors, and the resolution limit of the distance is effectively decreased when the distance is below the classical diffraction limits.

Nowadays, the photon numbers can be counted by new kinds of single-photon detectors. The circumstance that two photon are absorbed by one detector is a new exclusive event, as the new event contributes correlation terms, the usage of more precise detectors will further enhance the Fisher information. When the emitted photons from each emitter are in coherent state, the detecting intensity up to $n$-th order of multi-photon detection can be taken into consideration. Therefore, adding photon detectors can be a solution for resolving multi-photon lights. We hope our work can give a new vision on the relationship between super-resolution and distance measuring.

\begin{acknowledgments}
This work is funded by National Natural Science Foundation of China (No. 61275122, 61590932), the Innovation Funds and Strategic Priority Research Program (B) of CAS (No. XDB01030200), and the National Key R\&D Program (No. 2016YFA0301700).
\end{acknowledgments}

\end{document}